\documentclass[12pt]{article}
\date{}
\usepackage[T1]{fontenc}
\usepackage[utf8]{inputenc}
\usepackage{authblk}
\usepackage{hyperref}

\hypersetup{
    colorlinks=true,
    linkcolor=blue,
    }
\newcommand\blfootnote[1]{
  \begingroup
  \renewcommand\thefootnote{}\footnote{#1}
  \addtocounter{footnote}{-1}
  \endgroup
}

\title{Modelling the Dynamics of a Hypothetical Planet X by way of Gravitational N-body Simulator.}
\author[1,2]{Michael Cowley\thanks{michael.cowley@students.mq.edu.au}}
\author[3]{Stephen Hughes}
\affil[1]{Department of Physics \& Astronomy, 
Macquarie University, Sydney, NSW 2109, Australia}
\affil[2]{Australian Astronomical Observatory, 
PO Box 915, North Ryde, NSW 1670, Australia}
\affil[3]{Department of Chemistry, Physics and Mechanical Engineering, Queensland University of Technology, Gardens Point Campus, Brisbane, Queensland 4001, Australia}

\usepackage{graphicx}
\usepackage[font={footnotesize}]{caption}
\begin{document}

\begin{titlepage}
\noindent \large {This is an author-created, un-copyedited version of an article published in the European Journal of Physics. IOP Publishing Ltd is not responsible for any errors or omissions in this version of the manuscript or any version derived from it. The Version of Record is available online at \href{http://dx.doi.org/10.1088/1361-6404/aa5448}{http://dx.doi.org/10.1088/1361-6404/aa5448}}.
\end{titlepage}

\maketitle

\begin{abstract}

This paper describes a novel activity to model the dynamics of a Jupiter-mass, trans-Neptunian planet of a highly eccentric orbit. Despite a history rooted in modern astronomy, ``Planet X'', a hypothesised hidden planet lurking in our outer Solar System, has often been touted by conspiracy theorists as the cause of past mass extinction events on Earth, as well as other modern-day doomsday scenarios. Frequently dismissed as pseudoscience by astronomers, these stories continue to draw the attention of the public by provoking mass media coverage. Targeted at junior undergraduate levels, this activity allows students to debunk some of the myths surrounding Planet X by using simulation software to demonstrate that such a large-mass planet with extreme eccentricity would be unable to enter our Solar System unnoticed, let alone maintain a stable orbit. 

\end{abstract}

\section{Introduction}

The term ``Planet X" has a long history dating back to 1905, when the American astronomer, Percival Lowell, coined the term to describe an undiscovered ninth planet in our Solar System \cite{1}. Lowell came to this prediction after making observations of perturbations in Uranus and Neptune's orbital movement. 25 years later, at the Lowell Observatory in Arizona, Clyde Tombaugh used Lowell's predictions and made the discovery of Planet X. In maintaining the tradition of mythological naming, this distant planet was named Pluto after the Greek god of the dark underworld \cite{2}. With the discovery of Pluto, the term Planet X was freed for the potential discovery of other trans-Neptunian planets.

The appeal for undiscovered trans-Neptunian planets remains, and while large Kuiper Belt Objects, or “Dwarf Planets”, continue to be found \cite{3}, some astronomers still argue a much larger planet may exist \cite{4}. Unfortunately, the concept of Planet X has also been adopted by conspiracy theorists, many of whom have twisted it into doomsday scenarios around other frivolous claims. Despite lacking a sound basis in science or even cultural history, these stories have become a pop-culture phenomenon. A prominent contributor to these theories was Zechariah Sitchin, an Azerbaijani-born American author who claimed that his translation of ancient Sumerian texts led him to the discovery that the Sumerians knew of a large planet (known as Nibiru) beyond Pluto, which passes through our Solar System once every 3,600 years \cite{5}. This planet has been linked to a range of eschatological beliefs, including the 2012 phenomenon, which was the prediction of a sudden collision between the Earth and a large planet on or around the 21st of December 2012. While this date has since passed, the doomsday scenario remains and new dates for the planet's return, ranging from the present to AD 2900, continue to be touted \cite{6}.

\section{Method}

The activity described in this paper investigates the gravitational influence of a highly eccentric, Jupiter-mass planet on our inner Solar System, and its ability to maintain a stable orbit. We make use of the \emph{AstroGrav Astronomy Software} \cite{7}, which is an easy to use, graphical-based n-body simulator designed to educate students about celestial mechanics. Systems with multiple gravitating objects (stars, planets, moons etc.) are difficult to solve analytically, and an n-body simulator provides a fast and efficient alternative to study the dynamic evolution of these objects. Despite being education-based, \emph{AstroGrav} is built around a Verlet integration algorithm, and makes use of Newtonian Gravitation, Kepler's Laws of Planetary Motion and General Relativity to calculate orbital dynamics\blfootnote{The authors have no commercial interests in any software described in this article}. The software can be downloaded for a free trial, and includes a tutorial and additional support documentation\footnote{http://www.astrograv.co.uk/}; however, the activity described is not dependent on its use. Though not as user-friendly, numerous open source solutions are also available (e.g. Python-based \emph{PyEphem}\footnote{http://rhodesmill.org/pyephem/} and C++-based \emph{ORSA}\footnote{http://orsa.sourceforge.net/}). 

\subsection{Orbital Parameters}

To model our hypothetical Planet X, we first calculate its orbital parameters using Kepler's three laws of planetary motion:

\begin{enumerate}
  \item Planets move around the Sun in ellipses, with the Sun at one focus
  \item An imaginary line connecting the Sun to a planet sweeps out equal areas in equal intervals of time
  \item The square of the orbital period of a planet about the Sun is proportional to the cube of the semimajor axis of its orbit.
\end{enumerate}

\noindent By applying these laws to three assumptions made about our hypothetical Planet X (i.e. Jupiter-mass, long orbital period, and a collision with Earth), we can state:

\begin{enumerate}
  \item Planet X orbits around the Sun
  \item Planet X must move very slowly at aphelion (furthest point from the Sun) and very fast at perihelion (closest point to the Sun)
  \item Planet X's semi-major axis can be calculated using Kepler's third law (shown below)
\end{enumerate}

\enlargethispage{\baselineskip}

\noindent If the semi-major axis, $a$, is measured in astronomical units (1AU equals the average distance between the Earth and the Sun) and the period, $p$, is measured in years, then Kepler's third law can be expressed as $a^3 = p^2$.
\newline

By applying Newton’s Laws we find that Kepler’s third law takes a more general form:

\vspace{1 mm}
\begin{equation}{ a }^{ 3 }=\frac { G\left( { M }_{ 1  }+{ M }_{ 2 } \right)  }{ 4{ \pi  }^{ 2 } } { p }^{ 2 }\end{equation}
\vspace{1 mm}

\noindent where \emph{G} is the gravitational constant, and ${M}_{1}$ and ${M}_{2}$ are the masses of the two orbiting objects in solar masses (${M}_{\odot}$). In our example, ${M}_{1}$ and ${M}_{2}$ relate to the Sun and Planet X, respectively. Given the assumptions of Planet X (i.e. Jupiter-mass, $1{M}_{J} = 9.55 \times 10^{-4}$ ${M}_{\odot}$, and an orbital period of $p = 3,600$ years), solving for the semi-major axis yields $a =$ 234.9AU. For comparison, Pluto has a semi-major axis of 39.5AU and an orbit considered highly eccentric. As shown in Figure 1(a), the eccentricity, $e$, characterises how elongated an elliptical orbit is. 
\newline

\begin{figure}[ht]
\centering
\includegraphics[scale=0.25]{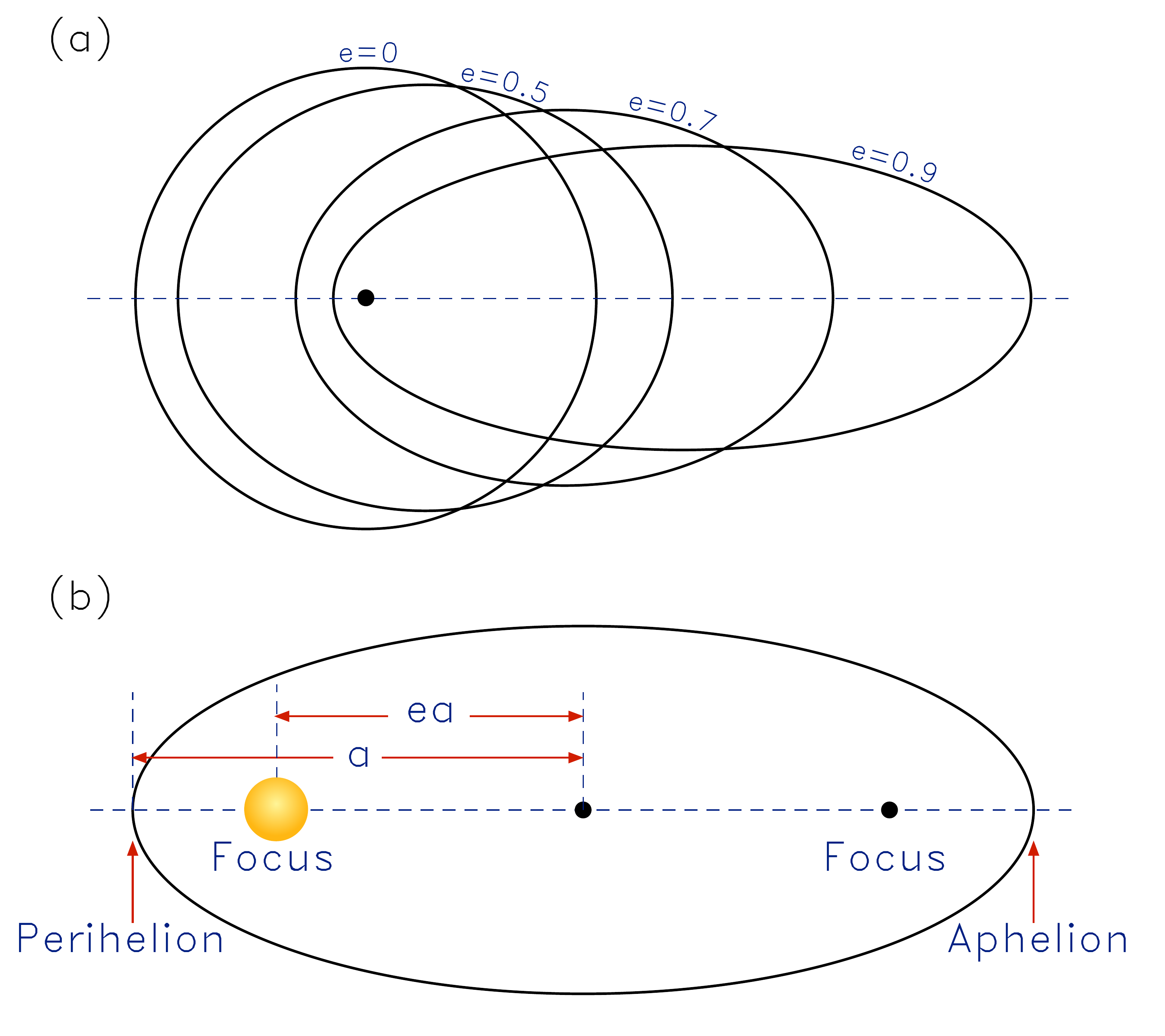}
\caption{Figure 1. (a) Elliptical orbits with an increasing eccentricity from $e=0$ (a circle) to $e=0.9$. (b) The geometry of an elliptical orbit showing the relationship between the focus, semi-major axis, $a$, and the eccentricity, $e$.}
\label{fig:fig1}
\end{figure}

Figure 1(b) illustrates how the eccentricity, $e$, the distance between the centre and focus, $f$, and the semi-major axis, $a$, are related to each other:

\begin{equation}f=ea\end{equation}

\noindent Assuming that perihelion for Planet X is at Earth's orbit (based on the assumption of a collision between the two planets), the distance between the centre of the orbit and the focus can be expressed as $f=a - 1\mathrm{AU}$. Rearranging equation (2), we can solve for the eccentricity, $e=f/a=\left( 234.9\mathrm{AU}-1\mathrm{AU} \right) /234.9\mathrm{AU}=0.996$, which suggests Planet X has a highly elongated elliptical orbit.
\newline

As stated above, Kepler's second law suggests the highly elongated orbit of Planet X results in a high orbital velocity at perihelion. We can quantify this amount using the orbital energy conservation equation (also known as the Vis-Viva equation), which relates kinetic energy to gravitational energy:

\begin{equation}E=\frac { 1 }{ 2 } m{ v }^{ 2 }-\frac { G{ M }_{ \odot  }{ M }_{ X } }{ r } =-\frac { G{ M }_{ \odot  }{ M }_{ X } }{ 2a } \end{equation}
\vspace{1 mm}
\begin{equation}
 v=\sqrt { G{ (M }_{ \odot  }+{ M }_{ X })\left( \frac { 2 }{ r } -\frac { 1 }{ a }  \right)  }\end{equation}

\noindent where \emph{r} is the distance between Planet X and the Sun (i.e. 1AU). Upon substituting our values into this equation, we yield a velocity of $v=42.1$km/s. 
\newline

From just three assumptions about Planet X: its mass, its orbital period, and its distance from the Sun at perihelion, we've been able to use established laws to determine numerous orbital parameters, which we can now use to model the dynamics of Planet X's orbit. However, even before completing any simulations, the stability of Planet X's orbit is brought into question due to the extreme values of its orbital parameters. For example, the escape velocity from our Solar System at a distance of 1AU is given by:

\begin{equation}{ v }_{ \mathrm{esc} }=\sqrt { 2G\frac { { M }_{ \odot  } }{ r }  } =42.2\mathrm{km/s}\end{equation}
 
For comparison, Earth's mean orbital velocity is 30 km/s, Venus's is 35 km/s and Mercury's is 47 km/s. It can be seen that at a distance of 1AU, Planet X would be orbiting at a similar velocity to Mercury's and a touch under its escape velocity. In the next section, we determine how susceptible this volatile orbit is to gravitational perturbations by modelling the dynamics of Planet X's orbit.

\subsection{Simulation Configuration}

Using the orbital parameters calculated in Section 2.1, we now model the dynamics of our hypothetical Planet X in our gravitational n-body simulator. While the primary goal of this simulation is to observe and record the interaction between Planet X and the planets of our Solar System, we take the novel approach of placing it in the context of a doomsday scenario - an impact with Earth on December 21, 2012. To achieve this, we start by loading an existing model of our Solar System in \emph{AstroGrav}, which includes the orbital parameters of all major planets. We then select the Sun as a parent and add a new planet in its equatorial plane (i.e. inclination, $i=0^{\circ}$) using the simulator parameters detailed in Table 1.

\begin{table}[h!]
  \centering
  \caption{Table 1. Simulator Parameters of Planet X.}
  \label{tab:table1}
	\begin{tabular}{l*{6}{c}r}
	Parameter			& Value \\
	\hline
	Mass (m)			& 1$M_J$  \\
	Semi-major Axis (a)	& 234.9AU  \\
	Eccentricity (e)	& 0.996  \\
	Period (p)			& 3,600 years
  \end{tabular}
\end{table}

By default, the simulator will place the planet at aphelion, the furthest point from the Sun. As a result, the simulation requires a start date of December 21, 212 AD, which is 1,800 years (i.e. $p/2$ years) prior to the doomsday scenario. Depending on computing power, the simulation time for such a scenario may be prohibitive. As an alternative, we provide optional orbital state vectors in Table 2, which are cartesian vectors of position and velocity (see Figure 2) that reposition Planet X's trajectory to $\sim$20 years before it reaches perihelion.

\begin{table}[h!]
\centering
  \caption{Table 2. Optional State Vectors of Planet X.}
  \label{tab:table2}
	\begin{tabular}{l*{6}{c}r}
	Parameter			& Value \\
	\hline
	Orbital Position x	& 4.60AU  \\
	Orbital Position y	& -39.9AU  \\
	Orbital Position z	& 0.00AU \\
	Orbital Velocity v$_\mathrm{x}$	& 0.30 km/s  \\
	Orbital Velocity v$_\mathrm{y}$	& 6.35 km/s\\
	Orbital Velocity v$_\mathrm{z}$	& 0.00 km/s
  \end{tabular}
\end{table}

\begin{figure}[ht]
\centering
\includegraphics[scale=0.35]{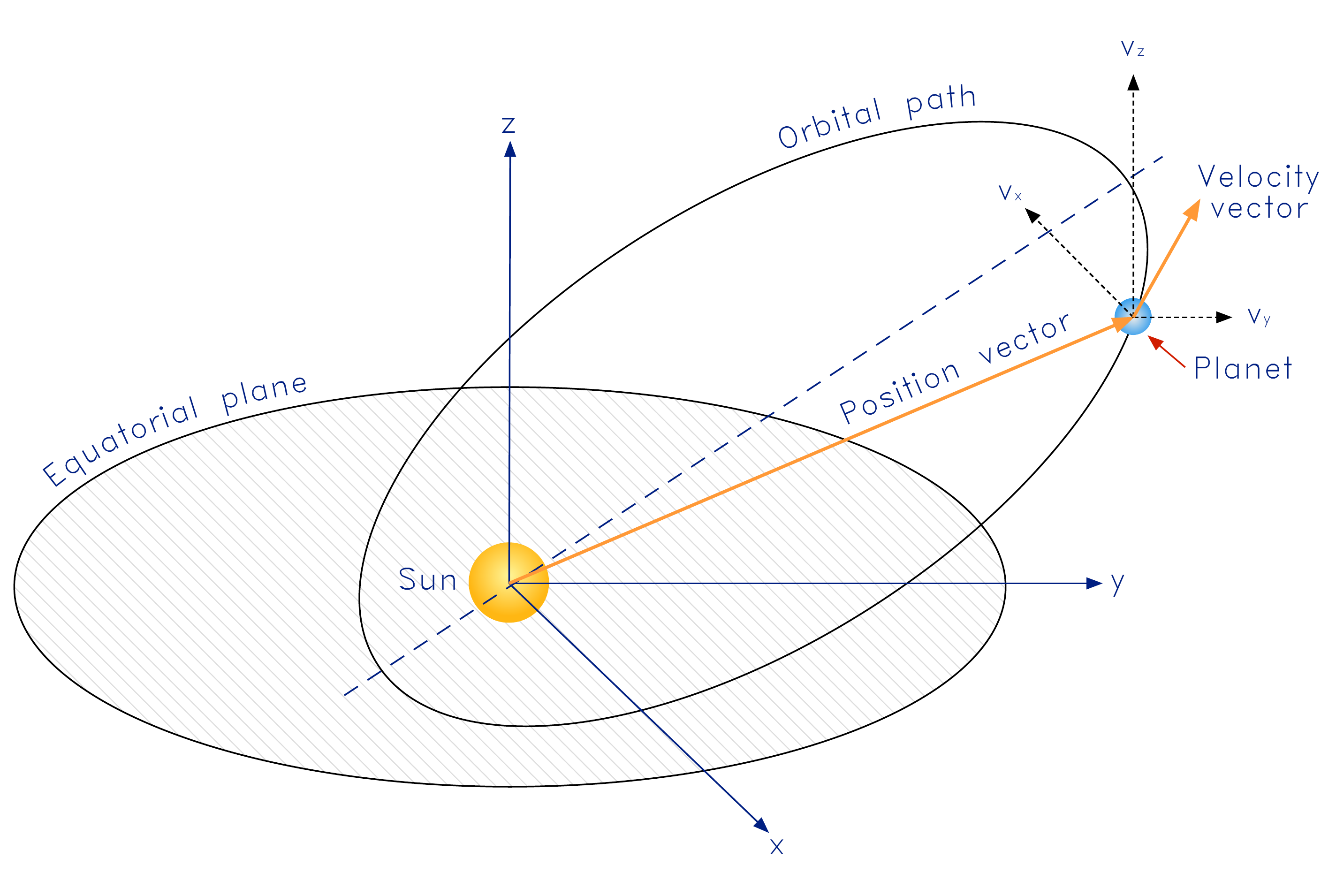}
\caption{Figure 2. Orbital state vectors. The position vector describes the position of the planet, while the velocity vector describes its velocity. Together, these state vectors describe the planet's trajectory.}
\label{fig:fig2}
\end{figure}

\enlargethispage{\baselineskip}

Despite repositioning the trajectory of Planet X, these state vectors have been calculated to retain the orbital parameters from Table 1. As shown in Figure 3, this will place Planet X just outside of our Solar System. To use these optional vectors, the simulator should be evolved to a start date of August 28, 1992 and the orbital properties of Planet X updated with those from Table 2. 

\begin{figure}[hb]
\centering
\includegraphics[scale=0.30]{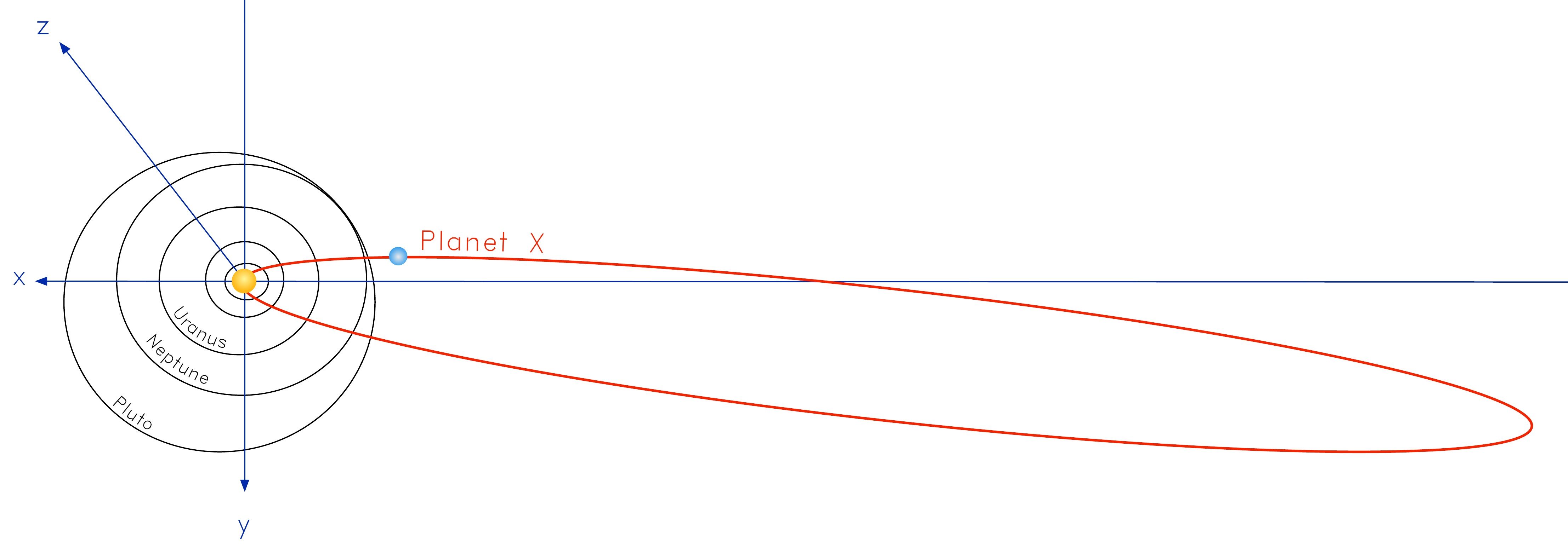}
\caption{Figure 3. The position and orbital path of Planet X using the optional state vectors in Table 2.}
\label{fig:fig3}
\end{figure}

\section{Simulation Analysis}

With the simulation configured, we now advance Planet X forward towards December 21, 2012, the collision date for our Planet X scenario. We use a time step of $dt=10$ days, which recalculates the velocity and position of Planet X every 10 simulated days. Months before this date, we note fluctuations of $\pm$days for the orbital periods of Earth and Mars. To ensure these fluctuations are not simulator errors, we repeat the simulation without Planet X. We find no evidence of these fluctuations, which suggests the previous results are driven by the gravitational influence of Planet X. 
\newline

We now allow the simulator to complete its run to December 21, 2012 and halt it just prior to Planet X's collision with the Earth. Data from the simulator was then exported as plain text and imported into the spreadsheeting software, \emph{Excel}, for formatting. We then compare the orbital periods of the terrestrial planets to the those from the simulation without Planet X, and present the data in Table 3.

\begin{table}[h!]
  \centering
  \caption{Table 3. Orbital Perturbations by Planet X.}
  \label{tab:table3}
	\begin{tabular}{l*{6}{c}r}
	Planet	& Orbital Period  	& Orbital Period		& Difference \\
	Name	& Unperturbed (days)	& Perturbed (days)	& (days)	\\
	\hline
	Mercury	& 87.87	& 87.98	& 0.11  \\
	Venus	& 224.7	& 225.2	& 0.30  \\
	Earth	& 365.3	& 397.0	& 31.7  \\
	Mars	& 687.0	& 683.9 & -3.10 \\

  \end{tabular}
\end{table}

As shown in the above table, Planet X had a substantial impact on the smaller terrestrial planets, perturbing their orbital periods by a significant number of days. For example, we find Earth's orbital period perturbed by 31.7 days, with a year now taking 397 days. As mentioned above, these perturbations were first recorded months before the collision date of December 21, 2012. Such extreme changes to our orbital period would most certainly be noticed, which suggests claims of a sudden and unexpected collision with Planet X is highly unlikely. 
\newline

\section{Discussion}

There is virtually an unlimited number of paths we could simulate for Planet X, which would result in an equally large number of outcomes. With that said, the eccentricity we calculate for Planet X suggests its orbit is highly unstable. Pluto, another highly eccentric planet, is currently in a 2:3 resonance with Neptune: for every two orbits that Pluto makes around the Sun, Neptune makes three. As a result, the two planets are highly stable and their orbits are expected to be preserved for millions of years to come \cite{8}. Unless a similar relationship exists between Planet X and another planet, the long-term stability of its orbit is highly questionable.
\newline

To test this, we reconfigure the simulation to allow Planet X to complete its orbit past perihelion. To achieve this, the existing simulation needs only to be adjusted with a random start date. This will ensure Planet X no longer collides with Earth. In our simulation, we adjusted the start date by one day and observed Planet X being ejected from the Solar System. Further variations to this start date resulted either in a similar fate for Planet X or at least in perturbations of hundreds of years to its orbital period. 
\newline

The results from our simulations highlight the chaotic behaviour of a large-mass planet with extreme orbital dynamics when released into our Solar System. Planet X not only significantly perturbed the orbits of the terrestrial planets, but failed to maintain a stable orbit of its own. 

\subsection{N-body Test}
It should be noted that n-body simulations do diverge from exact solutions due to round-off errors in the computer, and truncation errors caused by representing the continuous evolution of time as a discrete value. While we find no evidence arising from round-off errors, truncation in time step has the potential to significantly influence our results. For example, Mercury takes $\sim$88 days to orbit the Sun, and given the adopted time step of $dt=10$ days, the simulator calculated a new position for Mercury $\sim$9 times during its complete orbit. While this proved to be sufficient to preserve the orbit of Mercury in this simulation, (as shown in Table 3), an increase in time step may lead to orbital decay and instability. The result often depends on the integration method adopted by the simulator software. To test how the Verlet integration algorithm in $Astrograv$ handles large time steps, we place a Mercury-like planet in a circular orbit about a Sun-like star. We choose parameters to produce a circular orbit with a period of 88 days. For the first test, we set the integrator time step to $dt=10$ days, the same as our Planet X simulation, and evolve the simulation by 10,000 days. The result is shown in Figure 4(a). For the second test, we set the simulator to a larger time step of $dt=44$ days (half of the orbital period of Mercury) and again evolve it forward by 10,000 days. The result is shown in Figure 4(b). In both cases, the period of the orbit is determined more accurately by the simulator by interpolating the Keplerian solution.

\begin{figure}[h]
\centering
\includegraphics[scale=0.19]{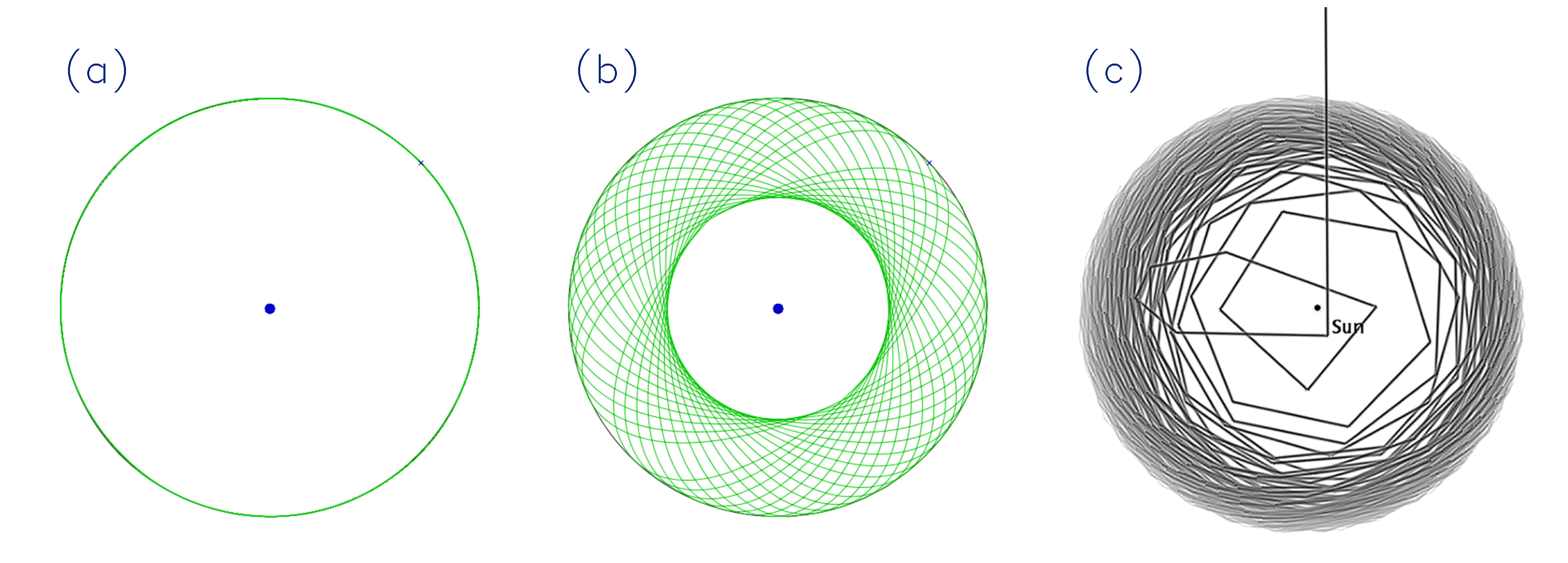}
\caption{Figure 4. A simulation of a Mercury-like planet in a circular orbit about a Sun-like star using (a) Verlet integration with a time step of $dt=$ 10 days, (b) Verlet integration with $dt=$ 44 days, and (c) Runge-Kutta integration with $dt=$ 10 days.}
\label{fig:fig4}
\end{figure}

As shown in the above figure, increasing the time step inhibits the ability to maintain a circular orbit. However, despite an increase in time step to $dt=44$ days, the simulator still produces a stable, yet precessing orbit with no sign of decay. For comparison, we conduct the same test using a Runge-Kutta integrator, but as shown in Figure 4(c), we find the orbit quickly decays. While this lends credit to the stability of the Verlet integration used here, we caution that a smaller time step, while taking longer, is preferential to minimise truncation errors.

\section{Conclusion}
In this paper, we have presented an activity that uses gravitational simulation software to demonstrate that a large-mass planet with extreme eccentricity would be unable to enter our inner Solar System undetected, let alone maintain a stable orbit. The activity takes a novel approach of placing the simulation in the context of a popular doomsday scenario - a collision between Earth and a large-mass planet. We capitalise on the appeal of this scenario to introduce students to celestial mechanics, while allowing them to debunk the pseudoscience surrounding this topical scientific issue.

An early version of this activity was performed at a university workshop for students, which also included an introductory session on the astronomy-related myths surrounding the 2012 phenomenon. Student reactions were positive, as they seemed not only to enjoy the workshop, but also learned something new. We also encourage the reader to review other novel approaches to calculate gravitational interactions between astronomical bodies \cite{9,10,11}.

\def\baselinestretch{0.90}

\section{References}

\begingroup
\renewcommand{\section}[2]{}

\endgroup

\end{document}